\documentclass[12pt]{article}

\usepackage{amsmath}
\usepackage{amsfonts}
\usepackage{amssymb}

\usepackage{amsmath}
\usepackage{graphicx}
\begin{document}
 
\newpage

\title{\bf{A Cyclical Baryonic Big Bang Explains the Universe}}

\author{David E. Rosenberg,\\  EVMS Norfolk VA. 23507 USA \\ }
\date{    }

\maketitle

\section{Abstract}
Our Universe has multiple examples of unexplained gravitational losses in black holes and neutron stars. The smallest black hole $\approx 4 M_\odot$ means the maximum baryonic density is $\rho \approx 10^{17}grams/cm^3$. Any collapse of the universe will stop with a scale factor $\approx 10^{13}$ cm. and radiation energy $\approx 10$ GeV. Due to higher squeezed core baryons, the outer part of the mass transferred heat energy to the core and became cold dark matter. After contraction reduced particle motion and gravitation, the core radiation energy propelled pieces of the shell into the universe. Each of these masses captured hot core gases according to its gravitational size, forming proto-galaxies. A cold shell and a hot core will explain the Planck spectrum and large galaxy formation in the early universe. Thus the universe was never radiation dominant. The universe will remain cyclical as any increase in the entropy of matter will be crushed back to neutrons during the contraction phase.

Key words: Cosmology theory, dark matter, dark energy, galaxy formation

\section{Introduction}
There are a number of unsolved problems with the current hot big bang model of the Universe. General relativity has been verified in the Universe except in extreme density conditions: the big bang and black holes. Galaxies are constructed similarly despite their origins being physically too fair apart to be in casual contact (the horizon problem). Initial spacetime was Minkowskian-the big bang expansion energy exactly matched the gravitational energy (the flatness problem). The massive pre-big bang matter must have been a black hole. There are large masses $9$ and  $10 \times 10^{10}M_\odot$, unexplained in the early Universe. The hot synthesis of light elements occurred in only $4-5\%$ of the matter present. Extrapolation of general relativity many orders of magnitude in the big bang and black holes to or near singularities has not been successful in solving the Universe problems. There is no evidence that the Universe ever reached Planck energies. In the 'Road to Relativity' Einstein felt the gravitational tensors were 'rock solid' but stress-energy tensors were made of 'wood'. The strange physics of highly squeezed matter will be explored here.  

Friedman-Robertson Walker geometry generates an isotropic homogeneous spacetime which prevents an initial big bang picture..
\begin{equation}
c^{2}d\tau=c^{2}dT-a(t)^{2}d\Sigma^{2}
\end{equation} where
\begin{equation}
d\Sigma^{2}=\frac{dr^{2}}{1-kr^{2}}
\end{equation}
where $k=-1,1,+1$ depending on whether the universe is open geometry, flat or closed.

Neutron stars are presumed to have crusts with a surface region densities  $< 10^{14} g/cm^{3}$. Here in beta equilibrium, there are electrons, neutrons, and nuclei. Relativistic degenerate electrons comprise most of the pressure. The baryon density is near nuclear saturation density $n_{0}\approx 0.16\,fm^{3}$. The general presumption has been that the core should be contain even more relativistic gravitational matter.  
Applying general relativity to a static spherical symmetric metric gives
\begin{equation}
ds^{2}=-e^{2\Phi(r)}c^{2}dt^{2}+e^{2\Lambda(r)}dr^{2}+r^{2}d\theta^{2}+r^{2}sin^(2)\theta\, d\phi^{2}
\end{equation}
where $\phi$ is the azimuthal angle, $\theta$ is the polar angle and the radial coordinate $r$ is defined such that  
at the origin the circumference of a circle is $2\pi r$. Using a Schwarzschild geometry at the star surface, the boundary condition mandates 
\begin{equation}
\Phi(r=R)=\frac{1}{2}ln \Bigl(1-\frac{2GM}{Rc^{2}}\Bigr)
\end{equation}
\begin{equation}
\Lambda(r=R)=-\frac{1}{2}ln\Bigl(1-\frac{2GM}{Rc^{2}}\Bigr)
\end{equation}
If one assumes a perfect fluid which oversimplifies calculations, this leads to the Tolman-Oppenheimer-Volkov (TOV) equations:
\begin{equation}
\frac{dP}{dr}=-\frac{Gm\rho}{r^{2}}\Bigl(1+\frac{P}{\rho c^{2}}\Bigr)\Bigl(1+\frac{4\pi r^{3}P}{mc^{2}}\Bigr) \Bigl(
1-\frac{2Gm}{rc^{2}}\Bigr)^{-1}
\end{equation}
\begin{equation}
\frac{dm}{dr}=4\pi r^{2} \rho
\end{equation}
Here $P=P(r)$, $\rho=\rho(r)$ and the mass within the radius $r$ is $m(r)$. Since the TOV equations are relativistic, the pressure $P$ will add to gravitation especially at the core. At this boundary condition,
$m(0)=0$, $\rho(0)=\rho_{c}$. With general relativity the pressure terms in the TOV equations will add much gravitation. However matter is not infinitely compressible.
It has been found that squeezing protons to 0.3 fm. (femtometer) yields an enormous resistive pressure of $10^{35}$ pascals\cite{Burkert et al. (2018)}. Squeezing protons to 0.1 fm will cause them to become highly repulsive. 
As they note, this requires more pressure than a neutron star core can generate. The following examples are going to show that gravitation is reduced both in neutron star cores and black holes.    

An unreasonable effectiveness of the post-Newtonian approximation has been found in strong gravitational fields of neutron stars\cite{Will (2011)}. The post Newtonian approximation assumes that gravitational fields in and around bodies are weak and the motions of matter are slow compared to the speed of light. Thus
\begin{equation}
(v/c)^{2}\sim GM/\tau c^{2} \sim p/\rho c^{2} << 0.1
\end{equation}
where $v, M \,and\, \tau$ are velocity, mass and separation of the system. Within masses, $p$ and $\rho$ are the pressure and density and $G \, and \, c$ are Newton's gravitational constant and the speed of light. The approximation in post-Newtonian calculation is the 'reduced' Einstein equation
\begin{equation}
(-\partial^{2}/\partial(ct)^{2}+\nabla^{2})h^{\alpha\beta}=-16\pi(G/c^{4})
\tau^{\alpha\beta}
\end{equation}
where $h^{\alpha\beta}$ is the deviation of the space-time metric $g_{\alpha\beta}$ from the flat Minkowski space-time metric $\eta_{\alpha\beta}$. They are related by the equation 
\begin{equation}
h^{\alpha\beta}\equiv\eta^{\alpha\beta}-(-g)^{1/2}g^{\alpha\beta}. 
\end{equation}
If gravity is weak here, then the gravitational tensor potential $h^{\alpha\beta}$ must be small and can approximate the  highly nonlinear equations of general relativity.

In 1974 a binary pulsar PSR \,$1913+16$ was observed. Both neutron stars had masses $\approx 1.4M_\odot$ in a quite relativistic orbital system with a mean speed $\approx 200 \, km/sec$. Unexpectedly, it was found that the rate of decay of the orbit was in agreement with the post-Newtonian quadrupole formula. More recently, the relativistic double pulsar $J \, 0737-3039$ performed according to the post-Newtonian calculations despite being in very strong gravitational fields. An effacement process had been postulated that the gravitational binding energy reduces the gravitational mass of each pulsar by $10-20\%$ compared to its rest mass.  

Another binary of low mass, AXJ1745.62901 has continuously been losing extra orbital momentum over a 20 year period\cite{Ponti et al. (2015)}. The high rate of orbital period decrease is $\dot{P}_{orb}= -4.03\pm 0.32 \times 10^{-11} s/s$. This is over an order of magnitude greater than expected loses due to gravitational waves, magnetic breaking, or conservative mass transfer. The path also included unexplained 'jitter' of $10-20$ seconds advancing or retarding the orbital period. Orbital loss explanations from accretion, a third body mass or an unrealistic $.001M_{\odot}$ in the outer disc were rejected.

Black holes will form at mass densities
\begin{equation}
\rho = (c^6)/(G^3M^2)
\end{equation}
where the constants $c^{6}/G^{3} = 6.272 \times 10^{84} grams^{3}/cm^{3}$. The smallest black holes found in today's Universe are $4M_{\odot}$ and will form at the highest squeezed baryonic matter $\approx 10^{17} grams/cm^{3}.$ Smaller black holes requiring higher densities to form are nonexistent. Best estimates of the total universe mass are $\approx 1.5 \times 10^{56}$ grams. After this collapsed, a mass with scale factor $\approx 10^{13}$ cm. will result with a radiation energy $\approx 10$ GeV. While this is enough energy to propel galaxies a good fraction of light speed, it is over a hundred tines too little to change baryons to a quark-gluon plasma. Highly squeezed baryons should require even more energy. 

A unexplained survivor of an encounter with our Milky Way Super-massive Black Hole has been found\cite{Bally (2015)}. From 2006 a G2 'cloud' was found accelerating toward our galactic center mass  $\approx 3\times 10^{6}\,M_\odot$. The closest approach took place about February 2014. By Sept. 2014 G2 cloud was mysteriously moving away from the black hole. It even failed to trigger a flare up of accretion activity. 

A strangely weak magnetic field of a $9 M_{\odot}$ black hole in the binary V404 Cygni system\cite{Dallilar et al. (2017)}. A burst of radiation was studied from a flare. It contained charged particles with electrons and protons in the black hole magnetic field. By calculating how quickly the burst dimmed, a team of astronomers found the magnetic field to be $461\pm 12$ gauss, the strength of several bar magnets and almost 3 orders of magnitude less than theory predicted.

A galaxy has been found which should not exist according to classical general relativity\cite{Bianchi (2019)}.  Galaxy NGC 3147 has a mildly relativistic Broad Line Region circling and very close $<100\,r_{*}$ to the central black hole in an $L/L_{Edel} \approx  10^{-4}$ system. This contradicts the current understanding of accretion flow configuration at extremely low accretion rates.  The only reason matter could be closely circling and not accreted in a massive galactic black hole is that the hole must have lost much of its gravitation.

\section{The Case Against Singularities}

Using general relativity, density infinities have been extrapolated in black holes and where the big bang started. 
If the universe started at or near a singularity and expanded, there should still be evidence of many of the following missing high energy phenomena. Monopoles are formed $>10^{14}$ GeV. Antimatter is formed $>38$ MeV. Domain walls would be the size of $10^{28}h^{-1}$ cm. and have a mass $\sim 4 \times 10^{65}\Lambda^{1/2}(\sigma/100 GeV)^{3}$ grams or many orders of magnitude over the present Hubble volume. Its presence would cause a large defect in the CBR.
Assuming the universe started as a fireball, the production of geometric flatness $\Bigl(a_{,t}/a\Bigr)^{2}=-k/a^{2}+\Lambda/3+\, 8\pi\rho_{m}a_{o}^{3}/3a^{3}+\,8\pi\rho_{r}a_{o}^{4}/3a^{4}$ with $k=0$ has required an unproven inflaton scalar $\phi$, leaving most other problems unsolved. Despite multiple attempts to find evidence of inflation, no evidence has been found in the CBR polarization or anywhere else. Fireballs can't make high correlations in galaxies with the velocity-brightness relation, corresponding rotations and all galactic parameters related to the central black hole size(see below).   
Considering the small size of the Higgs Boson (125 GeV), the big bang could not produce any cold dark nonbaryonic matter. Nucleosynthesis (the highest big bang energy confirmed) occurred after the end of quark-hadron boundary, requiring $\approx 95\%$ cold dark matter. 
 
\section{Explaining Gravitation Loses}
Gravitational masses have been treated as pinpoint sources. However, it is not logical that the gravitational properties of an infinitely collapsing highly squeezed mass would be the same as normal matter.  
For gravitational losses, matter must be highly squeezed and not a gas of noninteracting particles (perfect fluid) nor a Quark-Gluon Plasma. Nucleons highly squeezed to $0.3$ fm. must be packed so tightly by $10^{17}gm/cm^{3}$ that there is no more remaining space nor motion except quantum jitters at all.
In order to construct a stress-energy tensor \cite{Misner et al. (1973)}, the matter 4-velocity $\mathbf{u}=(dt,0,0,0)$, possibly adding only the observer 4-velocity. The 4-momentum is
\begin{equation}
\mathbf{p} = m\mathbf{u}=(m\gamma, m\Delta x \gamma, m\Delta y \gamma, m\Delta z \gamma)  
\end{equation}
where $E=m\gamma$, $\gamma =1$ nonrelativistic and $\Delta x, \Delta y, \Delta z =0$. A volume element $\Sigma$ composed of basis vectors $e_{x},e_{y}, e_{z}$ have zero magnitudes as there is no flow of 4-momentum.   
An observer in his Lorentz frame will measure mass-energy density in $gm/cm^{3} \, T_{00}=T(e_{0},e_{0})$, with the observers 4-velocity $\mathbf{u}$ replaced by $e_{0}$.  
If the box is at rest in the observers frame, all matter kinetic and quantum energy will be sufficiently damped except possibly quantum spin. Space-time interaction will cease, $T_{00}=0$.
To construct a volume in spacetime with a parallelopipid, use four different vectors for edges $\mathbf{A,B,C,D}$. The vectors in standard Lorentz frame are $\mathbf{A}=(\Delta t, 0, 0, 0)$, $\mathbf{B}=(0, \Delta x, 0, 0)$, $\mathbf{C}=(0, 0, \Delta y, 0)$ and $\mathbf{D}=(0, 0, 0, \Delta z)$. 
A 4-volume is
\begin{equation}
\Omega=\epsilon_{\alpha \beta \gamma \delta}A^{\alpha}B^{\beta}C^{\gamma}D^{\delta}=\mathbf{A} \wedge \mathbf{B} \wedge \mathbf{C} \wedge \mathbf{D}
\end{equation}
A volume integral of a tensor \textbf{S} defined over a four dimensional  region $\mathcal{V}$ of spacetime, calculated in a Lorentz frame 
\begin{equation}
M^{\alpha}_{\beta \alpha}=\int S^{\alpha}_{\beta \gamma} dt\, dx \, dy \, dz
\end{equation} 
 The energy density measured in such a volume $E=m\gamma/V=0$  as is the density of the 4-momentum $d\mathbf{p}/dV=0$ per 3 dimensional volume in an observers Lorentz frame.  In the following, $j,k=(1,2, or \,3)$ in what really is a symmetric tensor. $T^{j,0}=0$ is the momentum density, j component. $T^{0,k}=0$ is the energy flux, k component. $T^{j,k}=0$ is the j component of force from matter and fields acting around $x^{k}$.
This keeps the tensor divergence $\mathbf{\nabla} \cdot \mathbf{T}=0$ as there is no particle movement.

With rotating immobile squeezed nucleons, let S be a spacelike hypersurface with arbitrary event $\mathcal{A}$ and coordinates $x^{\alpha}(\mathcal{A})\equiv a^{\alpha}$ using globally inertial coordinates.  
Total angular momentum  on S about $\mathcal{A}$ can be defined as 
\begin{equation}
J^{\mu\nu} \equiv \int_{S} \mathcal{J}^{\mu\nu\alpha} d^{3}\sum_{\alpha}
\end{equation} and will add to total momentum only if present. 
Here 
\begin{equation}
\mathcal{J}^{ \mu\nu\alpha}\equiv(x^{\mu}-a^{\mu})T^{\nu\alpha}-(x^{\nu}-\mathcal{A}^{\nu})T^{\mu\alpha}
\end{equation}
If S is a hypersurface of constant time t then 
\begin{equation}
J^{\mu\nu}=\int \mathcal{J}^{\mu\nu 0} dx\, dy\, dz
\end{equation}
In the systems rest frame,  let $P^{0}=M$, $P^{j}=0$ and at the center of mass 
\begin{equation}
x_{cm}^{j}=\frac{1}{M}\int x^{j}T^{00}d^{3}x.
\end{equation}
For a large mass, intrinsic angular momentum  is defined angular momentum about any event $(a^{0},x_{cm}^{j})$ on the world line of the center of mass. 
Here components $S^{0j}=0$ and  
\begin{equation}
 S^{jk}=\epsilon^{jkl}S^{l}.\mbox{ and }  S\equiv \int (x-x_{cm})\times d\mathbf{p}/dV \,S^{\mu\nu} d^{3}x
\end{equation} 
The intrinsic angular momentum 4-vector $S^{\mu}$ has components in the rest frame (0,S) 
\begin{equation}
S^{\mu\nu}=U_{\alpha}S_{\beta}\epsilon^{\alpha\beta\mu\nu}
\end{equation} 
The 4-velocity center of a large highly squeezed mass $\mathbf{U}_{\beta}\equiv \mathbf{P}_{\beta}/M =0$.   
Angular momentum is composed of intrinsic and orbital parts. 
An arbitrary event $a$ whose perpendicular distance from the center of mass world line is $-Y^{\alpha}$ making $\mathbf{U}_{\beta}Y^{\beta}=0$.
The total angular momentum $J^{\mu\nu}$ about $\mathcal{A}$ is both the intrinsic part 
\begin{equation}
(S^{\mu\nu}=\mathbf{U}_{\alpha}\mathbf{S}_{\beta}\epsilon^{\alpha\beta\mu\nu})
\end{equation}
 and the orbital part 
\begin{equation}
(L^{\mu\nu}=Y^{\mu} P^{\nu}-Y^{\nu} P^{\mu})
\end{equation}
  With the angular momentum about $\mathcal{A}$ and the 4-momentum known (zero is this case) one can calculate
the vector from $\mathcal{A}$ to the center of mass world line.
\begin{equation}
Y^{\mu}=-J^{\mu\nu}P_{\nu}/M^{2}
\end{equation}
 
Using a swarm of identical particles with event $\mathcal{P}$ inside the swarm, $m_{A}$ is the rest mass. $\mathbf{u}_{A}$ is the 4-velocity, and $\mathbf{p}_{A}$ is the 4-momentum. 
$N_{A}$ is the number of particles per unit volume, as measured in the particles own rest frame. 
The number flux vector 
\begin{equation}
\mathbf{S}_{A}\equiv N_{A}\mathbf{u}_{A}
\end{equation}
 The particles have ordinary velocity $v_{A}$, zero in packed supranuclear densities. 
$\mathbf{u}^{o}_{A}$  is the the Lorentz correction for volume and velocity  $1/(1-v_{A})^{1/2}$. 
The 4-momentum density  is 
\begin{equation}
\mathbf{p}_{A} S^{o}_{A}=m_{A}u^{u}N_{A}u^{o}_{A}
\end{equation} 
Consequently the 4-momentum density has components 
\begin{equation}
T^{uo}_{A}= p^{u}_{A}S^{o}_{A}=m_{A}N_{A}u^{u}_{A}u^{o}_{A}
\end{equation} 
The flux of the $\mu$ component of momentum with perpendicular projection $e_{j}$ is 
\begin{equation}
 T^{\mu j}_{A}= p^{\mu}_{A}S^{j}_{A}=   m_{A}N_{A}u^{\mu}_{A}u^{j}_{A}
\end{equation} 
Here superscripts $(\mu,o)$ and $(\mu,j)$ of the frame independent equation
\begin{equation}
\mathbf{T}_{A}= m_{A}N_{A}\mathbf{u}_{A} \otimes \mathbf{u}_{A}= \mathbf{p}_{A} \otimes \mathbf{S}_{A}
\end{equation}
By summing over all categories, the total number flux vector and stress energy tensor are obtained for all particles in the swarm. If $\mathbf{u}_{A} =0$, these will be zero. 
\begin{equation}
\mathbf{S}=\sum_{A}N_{A}\mathbf{u}_{A} \mbox{ and }\mathbf{T}=\sum_{A} m_{A}N_{A}\mathbf{u}_{A} \otimes\mathbf{u}_{A}=\sum_{A}\mathbf{p}_{A}\otimes\mathbf{S}_{A}
\end{equation} 
The total momentum flux accross a closed 3-dimensional surface must vanish $\oint T^{\mu 0} d^{3}\sigma_{a} =0$. 
There is no flux and no sinks and there is no momentum at these supranuclear densities. 

Thorne has a conjecture that a black hole may form only when a given amount of mass-energy collapses through its own Schwarzschild radius $R_{S}=2M$, thus achieving compactness $M/R>0.5$\cite{Thorne (1970)}. A sufficient large collapsing shell is necessary. If one starts with collapsing black hole nucleons, each weighing $\approx 1.672\times 10^{-24}$ gm., maximally squeezed to $0.3$ fm radius and each with internal repulsive pressures $\sim 10^{35}$ pascals, they can not be packed to more than $\approx 10^{16}gm/cm^{3}$ density. Due to complete core gravitational loss with thermal jitters and possibly zero point motion suppressed (use stress-energy tensor formation above), collapsing matter can not overcome these internal pressures. This is the core of black holes. Not near Planck energies. Like neutron star cores, black holes will begin to lose gravitation by $\approx 5 \times 10^{14} gm/cm^{3}$. Once the core begins to form, its little to no gravitational energy and virtual non-compressibility blocks further gravitational collapse. It takes a total of $4M_{\odot}$, including the non-gravitational core to produce a black hole. The presence of a nongravitational core will reduce $M_{gravitation}$ but otherwise not affect gravitational waves in black hole-black hole coalescence. 

Gravitational losses in neutron stars must begin in their cores with density ranges $10^{14-15} gm/cm^{3}$. With a normal stressed medium, thermal jitters suppressed,  having velocities $|\mathbf{v}|<< 1$ with respect to a specific Lorentz frame, the spacial components of the momentum are $T^{0j}=\sum m^{jk} v^{k} $. Here 
\begin{equation}
m^{jk}= T^{\bar{0}\bar{0}}\delta^{jk}+ T^{\bar{j}\bar{k}}
\end{equation} 
Here $T^{\bar{\mu}\bar{\nu}}$ are components of the stress energy tensor in the rest frame of the medium. 
Inside a neutron star core, where velocities are known low, and probably zero $T^{\bar{0}\bar{0}} =0 \sim T^{\bar{j}\bar{k}}$. 

How to produce dark matter. After a gravitational collapse is halted by high squeezing of neutrons, the most highly squeezed particles in the core will act as an energy sink. 
Geometrized units will be employed such that $c=1=G$. 
From the second law of thermodynamics
\begin{equation}
T\, ds=d(\rho V) + pdV=d[V(\rho + p)]-Vdp
\end{equation}
here $s$ is the entropy of the matter and $V \propto a^{3}$ is the co-moving volume. 
Integrating gives $dp=(\rho + p)/T \, dT$. Substituting this into the above gives two equations,
\begin{equation}
d/dt\Bigl[(V(\rho + p))/(T)\Bigr]=(VE_{s})/(T)
\end{equation} 
and
\begin{equation}
\int ds= \int d\Bigl[ (V(\rho + p))/(T) \Bigr]
\end{equation}
The shell entropy change is $\dot{s}_{2}=E_{s}V/T_{2}$ and the core is $\dot{s_{1}}= - E_{s}V/T_{1}$.
Normally the energy and entropy increment would follow the temperature differential to a lower temperature as
with radiation and dust or with a scalar field and radiation.
\begin{equation}
\dot{s}_{total}=\dot{s_{2}} + \dot{s_{1}}=E_{s}V\Bigl(1/T_{2} -1/T_{1} \Bigr)
\end{equation}
Quantum effects cause some strange properties. 
Temperature is a function 
of the velocity or kinetic energies of the particles, $T^{o}K=m\bar{v}^{2}/3$. 
\begin {equation}
\int d\rho=\int ((\rho+p)/(n)\,dn-nTds)
\end{equation}
Due to more restricted motion of core baryons, shell energy will be transferred to the core. Core particles will have increased potential energy of deformation but further restricted movement reducing entropy and kinetic energy. 
Thus cold dark matter originated in shell baryons. The big bang hot core was about $4-5\%$ of the total mass.
There is still baryon conservation $dn/d\tau =-n\nabla \cdot u$. Energy conservation will still occur.
Only $d(nsV)/d\tau \leq 0$. Quantum gravity effects will put matter in better order, that is with less entropy and kinetic energy. 

Two teams of astronomers published work in $1997-1998$ trying to determine the geometry of the Universe by using Supernova Type 1A as standard candles\cite{Perlmutter (1997)},\cite{Riess (1998)}. They found that the supernova were $10-15\% $ further away than even a low density Universe $\Omega_M = 0.2$. Some negative gravitation or dark energy was canceling all the universe gravitational mass-energy. 
A black hole energy loss will affect the measured supernova distances of co-moving galaxies, as shown in Fig.~1. 
\begin{figure}
\setlength{\fboxrule}{1pt}	
\includegraphics[width=8.in]{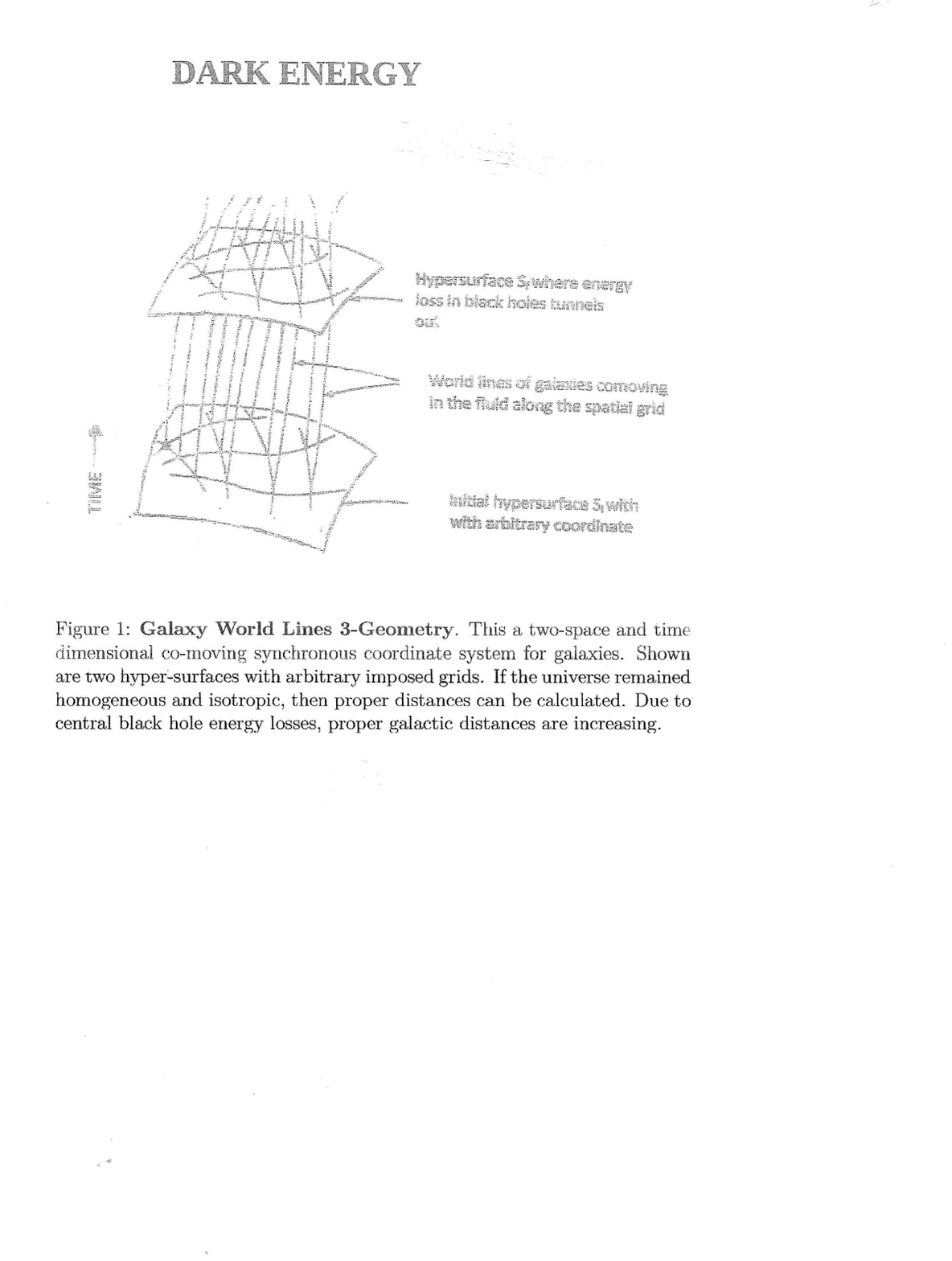}
\caption{\textbf{Galaxy World Lines 3-Geometry}. .}
\end{figure}
Spacetime geometry $d\sigma^2= g_{ij}(t,x^t)dx^idx^j$ of each hyperspace is assumed to be the same
due to homogeneity of the universe. The initial hyper-surface $S_I$ 3-geometry is $\gamma_{ij}x^K \equiv g_{ij}(t_I,x^K)$. At time $t_I$ on surface $S_I$, they are separated by
the proper distance 
\begin{equation}
\Delta\sigma(t_I)=(\gamma_i,\Delta x^i \Delta x^j)^{1/2}. 
\end{equation}
At some later time                        
$t_f$, they will be separated by some other proper distance $\Delta\sigma(t_f)$. When spacetime is
isotropic, then the ratio $\Delta\sigma(t_f)/\Delta\sigma_(t_I)$ will be related to the Universe scale factor $a(t_f)$ at time $t_f$. 
The Wilkinson Microwave Anisotropy probe combined with the Hubble Space Telescope resulted
in a very small value for the cosmological constant $\Lambda=3.73\times 10^{-56}cm^{-2}$ which corresponds to 
$\Omega_{\Lambda}=(\Lambda)/(3H_{0}^{2})=0.721\pm .015$\cite{Komatsu et al. (2011)}. Here $H_{0}=70.1 \pm 1.3 \, km/sec/Mpc$.
The loss in gravitational energy will cause galaxies no longer to be co-moving and to move away from the Hubble flow. The increasing distances measured between them is known as dark energy. 

\section{A Cyclical Universe} 
 The Friedmann equation postulates a perfect fluid to start. It will be valid only hours into the big bang when large shell fragments captured hot core gases forming proto-galaxies.  
The equations describing the scale factor evolution originate in the Ricci tensor. The 0-0 component of the Einstein equation is the Friedmann equation needing energy changes. In a Friedmann universe, radiation density is inversely related to the fourth power of 
the scale factor $\rho_{r} \propto a^{-4}$ and 
matter follows the third power $\rho_{m} \propto a^{-3}$.  
\begin{equation}
\Bigl(a_{,t}/a\Bigr)^{2}=-k/a^{2}+\Lambda/3+\, 8\pi\rho_{m}\,a_{o}^{3}/3a^{3}+\,8\pi\rho_{r}\,a_{o}^{4}/3a^{4}
\end{equation}
Here $a_{o}$ is our present day universe and $\Lambda$ is Einstein's term for energy of empty space. The question has long been why the initial geometry of the universe was flat or $k=0$ in  
the above equation. With the limiting matter density $\rho_{m}\approx 10^{17}grams/cm^{3}$, the universe was never radiation dominant. If all the initial radiation $\rho_{r} $ was embedded in the matter and participated in the subsequent expansion and gravitation of a fairly large mass $r_{radius}\approx 10^{13}$ cm., then the flatness problem is explained. Due to the matching of gravitation and expansion energies ($k=0$ above), it is most unlikely anything but nucleons stopped the prior universe collapse or restarted the expansion. There was no free radiation prior to the big bang. The Planck spectrum core photons participated in the re-expansion and nucleosynthesis. The small dark energy value attributed to $\Lambda$ is actually due to later galactic black hole gravitational loss as explained.
There is a question of increases of entropy on repeated universe cycles as irreversible physical processes may cause increasing cycles
\cite{Tolman (1931)}. 
\begin{equation}
\frac{dS_0}{dt}=\frac{1}{T_0}\frac{dE_0}{dt}+\frac{p_0}{T_0}\frac{d(\delta v_0)}{dt}+ \frac{\partial S_0}{\partial N_1}\frac{dN_1}{dt}+
...+\frac{\partial S_0}{\partial N_n}\frac{dN_n}{dt}
\end{equation}
As this is an adiabatic process in our universe, any entropy changes would be restricted to $N_1\cdots N_n$. Since the universe goes through a black hole where only mass, electric charge and angular momentum count, any entropy gain would be lost when the highly squeezed nucleons are produced.  
Our universe cycles as follows as shown in Fig.~2.
\begin{figure}
\setlength{\fboxrule}{1pt}
\includegraphics[width=5.in]{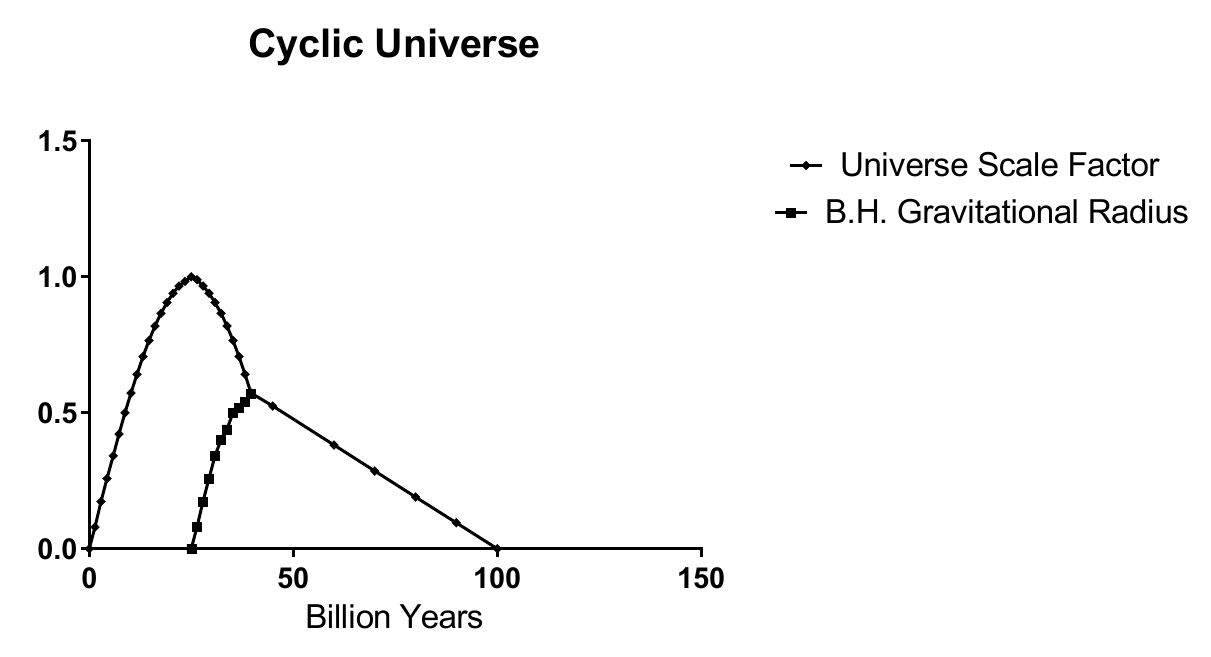}
\caption{\textbf{The Cyclical Universe} started from a spherical shaped mass at or near limiting density. Due to the bounce, the Universe expands to a maximum and then contracts. During contraction, there is a growing ultra-max black hole.  The collapsing universe will eventually force all matter and radiation into its gravitational radius. Kinetic energy of shell particles and radiation will slowly transfer to core. The gravitational energy and gravitational radius will slowly decrease to zero (cycle time estimated). Then core potential energy can start a new cycle of same size. Entropy is not increased.}
\end{figure}
During the universe contraction phase, there is a growing ultra-max black hole. All matter and photons inside its growing gravitational radius will follow null geodesics into the inside matter. Their energy will be slowly transferred to highly squeezed core particles. The collapsing universe will eventually force all remaining matter and radiation into the growing $r_{+}$. With the loss of gravitational energy and entropy, a subsequent universe bounce will not increase in size from the previous cycle. By bounce time, the gravitational radius has decreased to zero, $r_{+}\rightarrow 0$.
The very energetic core nucleons, loaded with radiation, powered the re-expansion as well as resulting gravity and nucleosynthesis.  
Nucleosynthesis thus occurred only in the core, leaving the shell as cold dark matter. 
If all the matter in the universe were in one super-mass, its radius would be $\sim 10^{13} cm.$ Its $4-5\%$ core released photons with an initial temperature $\sim 10^{13}$ degrees in a Planck spectrum. 
The cold baryonic shell surrounding the hot core absorbed and did not reflect hot core photons. 
It comprised a cavity close to the characteristic of a perfect black body with resulting 
radiation in thermal equilibrium.

Ihe initial mass expansion consisted of a cold shell surrounding the hot core. After thermal energy had been transferred to highly confined core baryons producing a Planck spectrum, gravitation loss permitted the expansion. The shell originated dark matter and black holes.  Whether released from a small hole or the massive break 
up of the shell, subsequent light emission will have a Planck spectrum with temperature fluctuations $\sim 10^{-5}$. 
The energy density was
\begin{equation}
u_\nu (T)= \Bigr( 8\pi\nu^2/c^3\Bigl) \Bigr(h\nu/(e^{h\nu/k_BT}-1)\Bigl)
\end{equation}
where $k_B$ is Boltzmann's constant. The first term on the right represents the number of electromagnetic modes of the standing waves at frequency $\nu$ per volume of cavity. The second term represents the average energy per mode at this frequency. The primordial spectrum of curvature perturbations can be represented as a power spectrum
\begin{equation}
\mathcal{P}(k) \propto k^{n_s -1}
\end{equation}
WMAP\cite{Komatsu et al. (2011)} showed that the power spectrum $n_{s}= 0.963 \pm .014$, which is nearly scale free.
The heights of the acoustic peaks are related to the related to the densities of the hot
and cold (CDM) baryons. The bulk modulus is reduced by increasing baryon fraction which
adds inertia but not pressure to the plasma. In the middle of the oscillations, the over-density
increases the compression peaks $(1,3,5 \ldots \,\,)$. The measured $\Omega_{b}h^{2}=0.02258 \pm .00056$
is consistent with nucleosynthesis. The angular scale of the acoustic peaks are related
to $r_{s}/d_{a}$, which is the sound horizon co-moving distance divided by the angular distance back to
the last scattering. This allows the spacetime curvature from the first acoustic peak to be measured. 
It was found to be flat to within $0.5\%$, all consistent with a bounce. 

\section{Baryonic Galaxy Formation}

Big bang fireballs can not originate highly structured galaxies\cite{Cattaneo et al. (2009)},\cite{Kormendy et al. (2011)},\cite{Lerner (2017)},\cite{Peebles (2011)},\cite{Rosenberg (2020)}. 
Galactic properties have been shown to be a function of only one variable: the central black hole mass\cite{Vandenbergh (2008)}. For most galaxies, there is a uniform history for galactic evolution\cite{Steinhardt (2014)}. There is a synchronization timescale $(\tau_{s}\approx 1.5 Gyr)$ where  galaxies of fixed mass and red-shift go through a deterministic sequence of star formation, quasar accretion and eventual quiescence. This sequence negates the importance of stochastic processes.
Galaxies can not form in a radiation dominant era so the very early galaxies found by the James Webb telescope are unexplained. The integrated Sachs-Wolf effect 
\begin{equation}
\ell^{2}C^{ISW}_{\ell} \simeq 72\pi^{2}/25\ell\int^{r_{LS}}_{0} dr\, r
\, g^{\prime 2}(r) \mathcal{P}_{\mathcal{R}}\Bigl(\ell/r\Bigr)T^{2}
\Bigl(\ell/r\Bigr)
\end{equation}
was designed for photons descending into and emerging from a gravitational well. Here $\mathcal{P}_{\mathcal{R}}(k)$ is the primordial co-moving curvature spectrum.
$r_{LS}$ is the co-moving radius to the last scattering surface. 
$g$ is the $\Lambda$ growth suppression factor. $T(k)$ is the transfer function for suppression during radiation domination. $k^{\prime}$ is the conformal time derivative of the co-moving wave number. 
Black hole masses greater $ 10^{4} M_\odot$ surrounded by hot gasses could have been present in the last scattering 
surface without enlarging the isotropy from $10^{-5}$. 
Starting from a big bang shell and hot core, early galaxy formation and high inter-galactic correlations become simplified. The shell laid down a fairly even cold dark matter density $\rho_{DM}$. 
The DM halo mass distribution for galactic systems ranging from dwarf discs
and spheroidals to spirals and ellipticals has been found essentially constant\cite{Donato et al. (2009)}. This result also spans almost the whole galaxy magnitude range $M_B$ from
$-8$ to $ -22$ and gaseous to stellar mass fraction range of many orders of magnitude.
\begin{equation}
log(\mu_{0D}/M_\odot pc^{-2})=\, 2.15 \pm 0.25
\end{equation}
where $\mu_{0D}$ is the central surface density and is defined as 
$r_0\rho_0$. $r_0$ is the halo core radius and $\rho_0$ is the central density. This same finding was supported by another group
\cite{Gentile et al. (2009)}.
Since all dark matter lies within a halo orbiting the primordial black holes, its total density field is
\begin{equation}
\rho(\mathbf{x})=\Sigma_{i} \int dm \int d^{3}x\prime \delta(m-m_{i})\delta(\mathbf{x}\prime-\mathbf{x}_{i})m\, u(\mathbf{x}-\mathbf{x}\prime|m)
\end{equation}
where $i$ is the different halos, $u=\rho/M$ is the normalized density profile and $M$  is the halo virial mass. 
Within the original dark matter halos, massive black holes coalesced. 
The dark matter from the center to periphery of early type galaxies has been evaluated from a galactic stellar mass $M_{*}\sim 10^{10}M_{\odot}$ to the more massive galaxies $10^{12}M_{\odot}$\cite{Tortora et al. (2014)}.
N-body simulations predict the the dark matter density profile $\rho_{DM}(r)$ should be independent of halo mass. In the NFW profile it is described by two power laws. In the outer regions it is $\rho_{DM}(r)\propto r^{-3}$ and in the center $\rho_{DM}(r)\propto r^{\alpha}$. Here $\alpha$ can vary $-1$ or $-1.5$ depending on the model. What has actually been found is a variation around the inverse gravitational square law ($\alpha=-2$), as the halos are orbiting the primordial black holes.
Galaxies larger than $ 3\times 10^{10}M_{*}$ have lower slopes than 
 $\alpha\approx 2$ due to accretion of halo mass from smaller satellite galaxies. Smaller satellite galaxies under $3 \times 10^{10}M_{*}$ have larger slopes due to loss of outlying halo matter.  
The rotation of the larger galaxies above this break point has been found  highly correlated and perpendicular to the filament that they are located
\cite{Dubois et al. (2014)}. The rotation of $65$ galactic black holes has been found aligned using their radio galactic jets \cite{Taylor and Jagannathan (2016)}.  

At first large and small shell masses were driven out forming filaments. Larger masses coalesced into black holes which held a fairly even density of smaller dark matter. Later hot core gas was captured according to the depth of the gravitational well. During capturing process, larger black holes, unlike smaller black holes, did not change the direction of rotation.
The deeper the gravitational 
wells, the higher the velocity and more orbiting mass that could 
be captured. 
The capturing process described is divided by distance from the primordial black holes $M$.
Outside the immediate area of black hole influence, capturing of hot core matter $m$ streaming through the area of influence of each black hole is due to the amount of energy each particle possesses. Large
kinetic energies result in hyperbolic or parabolic type orbits with the ability to escape any given gravitational well. Matter that is captured has the potential energy greater than the kinetic, 
\begin{equation}
GmM/r>\, l^2/mr^2\,+1/2 \,m\dot{r}^2
\end{equation}
and $e<1$. Expanding the total kinetic energy $E$ in the equation for $e$,
\begin{equation}
e=\Big[1+(2l^2(l^2/mr^2+\,1/2\, m\dot{r}^{2}-GmM/r))/mk^{2}\Big]^{0.5}
\end{equation}
Orbiting matter has $e<1$ and real. If we let its angular momentum $l=mr\dot \theta^2$ and $k=mMG$, the equation for $e$ becomes 
\begin{equation}
e=\Big[1+(r^6\dot{\theta}^{4}+\dot{r}^{2}r^{4}\dot{\theta}^2
-2GMr^{3}\dot{\theta}^{2})/(M^{2}G^{2}\Big]^{0.5}
\end{equation}
Using $\dot\theta=\dot r/r$, the equation for $e$ becomes
\begin{equation}
e=\Big[1+(2r^{2}\dot{r}^{4})/(M^2G^2)-(2r\dot{r}^{2})/(MG)\Big]^{0.5}
\end{equation}
As $GM=\dot r^2r$, 
then the galactic well will deepen as $M\propto \dot r^2$
or $M\propto r$. The last term in equation above becomes $\dot{r}^{8}/M^{2}G^{2}$. When this
term is dominant, it will allow capturing matter with $\dot{r}$ to 
increase as the fourth power as the galactic black hole $M$ increases,
$\dot{r} \propto M^{4}$. This explains the Tully-Fisher and similar correlations\cite{Shaya and Tully (2013)}.
The black hole capturing cross sectional area, $M_{csa}\propto M_{gravity}$ since both scale as $r^{2}$. 

The two stage gravitational formation process preserves angular momentum from the big bang.. Halo parameters are related to the luminous mass distribution since all rotating mass was captured by a given size black hole.  
An entirely baryonic model explains why the circular orbital speed 
from luminous matter, which dominates the inner regions, is so similar
to dark matter at larger radii. With many stars in the center areas, initial
conditions for dark and luminous matter no longer have to be closely adjusted 
to produce a flat rotation curve\cite{Ibata et al. (2013)}. 
The hot core matter of a certain velocity can be captured by the similarly sized black holes, 
explaining why there are similar circular speeds in 
all galaxies of a given luminosity no matter how the luminous matter is spaced. The depth of the gravitational well determines the circular speed
and luminosity of captured matter. The hot and cold matter discrepancies are detectable only at
accelerations below $\sim 10^{-8}cm/sec^{2}$ since they are all baryons. Much of the missing baryonic matter has been found in the intergalactic medium\cite{Nicastro (2018)}.

\section{Discussion}
General relativity has been extrapolated in black holes and the big bang to 
enormous energies and densities without consideration of properties of matter. Lightly squeezing nucleons like neutron stars will result in a light reduction of gravitational energy. Highly squeezing nucleons as in pre-big bang matter and black holes will result in elimination of gravitational energy.  
Highly squeezing matter will cause quantum effects in the stress-energy tensor so that $T_{\mu\nu}\rightarrow \mathbf{0}$. Very simple 'quantum gravity' can be produced. The gravitational losses including dark energy can't be explained any other way. Once damping of motion in highly squeezed particles is included, general relativity remains applicable throughout the Universe. A cold shell and hot core produced
the initial Planck spectrum radiation. The cold shell dispersion led to the great wall, filaments and voids.
The basis of almost all galaxies formed simultaneously as cold dark shell matter coalesced into black holes. which captured subsequent hot core gases into proto-galaxies. The formed stars ionized the intergalactic medium. The fact that the Higgs Boson is $125$ GeV and not larger, eliminates nonbaryonic matter as dark matter. 

Gravitation is described by the Riemann metric.  No negative energies are required nor is a vacuum energy necessary for empty space.
The fact that initial universe had flat space-time and a Planck spectrum has been explained. 
Since the origin of the big bang is a bounce, it should re-collapse into the big crunch. The resulting neutrons will eventually decompose into the protons and electrons to make hydrogen and helium for the next cycle.   
Black holes must continuously accrete mass-energy to maintain 
their gravitational strength. Gravity is not caused by matter itself but rather by the motion of matter particles.  Galaxies were made in a two stage process.  First the big bang shell laid down all halos and super-massive black holes. Then hot core gases were captured according to the size of each gravitational well.
\twocolumn

\bibliography{galaxy800.tex}

\end{document}